\pdfoutput=1
\documentclass[twocolumn]{aastex6}

\usepackage{graphicx} 
\usepackage{epstopdf}
\usepackage{ulem}

\shorttitle{A two-step confined flare}
\shortauthors{Zou et al.}

\begin{document}

\title{A Two-Step Magnetic Reconnection in a Confined X-class Flare in Solar Active Region 12673}

\author{Peng Zou\altaffilmark{1}, Chaowei Jiang\altaffilmark{1,2}, Xueshang Feng\altaffilmark{2,1}, Pingbing Zuo\altaffilmark{1}, Yi Wang\altaffilmark{1}, Fengsi Wei\altaffilmark{1}}

\altaffiltext{1}{Institute of Space Science and Applied Technology,
  Harbin Institute of Technology, Shenzhen 518055, China,
  chaowei@hit.edu.cn}

\altaffiltext{2}{SIGMA Weather Group, State Key Laboratory for Space
  Weather, National Space Science Center, Chinese Academy of Sciences,
  Beijing 100190, China}

\begin{abstract}
Solar flares are often associated with coronal eruptions, but there are confined ones without eruption, even for some X-class flares. How such large flares occurred and why they are confined are still not well understood. Here we studied a confined X2.2 flare in NOAA 12673 on 2017 September 6. It exhibits two episodes of flare brightening with rather complex, atypical ribbons. Based on topology analysis of extrapolated coronal magnetic field, we revealed that there is a two-step magnetic reconnection process during the flare. Prior to the flare, there is a magnetic flux rope (MFR) with one leg rooted in a rotating sunspot. Neighboring to the leg is a magnetic null-point structure. The sunspot drives the MFR to expand, pushing magnetic flux to the null point, and reconnection is first triggered there. The disturbance from the null-point reconnection triggers the second reconnection, i.e., a tether-cutting reconnection below the rope. However, these two reconnections failed to produce an eruption, because the rope is firmly held by its strapping flux. Furthermore, we compared this flare with an eruptive X9.3 flare in the same region with 2 hours later, which has a similar MFR configuration. The key difference between them is that, for the confined flare, the MFR is fully below the threshold of torus instability, while for the eruptive one, the MFR reaches entirely above the threshold. This study provides a good evidence supporting that reconnection alone may not be able to trigger eruption, rather, MHD instability plays a more important role.
\end{abstract}


\section{Introduction} \label{sec:intro}

Solar flares and coronal mass ejections (CMEs) are the most violent activities on the Sun. Now it is well recognized that they are different manifestations of the same process: sudden reconfiguration and energy release of the magnetic field in the solar corona. Observations show that flares and CMEs are closely related to each other~\citep{zhang01,zhang04,qiu04,temmer08}. Statistical studies reported that $\sim 90\%$ of X-class flares are associated with CMEs~\citep{yashiro05,wang07}. There are also CME-less X-class flares, and some of them are totally confined, that is, no eruption can be seen during the flares, which is different from flares in which coronal eruption can be seen but fails to escape into the interplanetary space \citep[e.g.,][]{ji03}. For example, in the super active region (AR) NOAA~12192, 6 X-class flares occurred with the largest one reaching X3.1, but none of them was associated with CME or coronal eruption. How such large flares occurred and why they are confined are still not well understood. 

Magnetic reconnection is the central mechanism in producing flares, and many theoretical models have been proposed to explain how the coronal magnetic field can be led to reconnect and/or erupt.
A major part of the models~\citep[e.g.,][]{torok05,kliem06,fan07} assume that prior to flare, there exists a coherent set of twisted magnetic flux, known as magnetic flux rope (MFR), which is subjected to some kind of ideal magnetohydrodynamics (MHD) instabilities and erupt during the flare, forming a CME. During the eruption, reconnection can then be triggered below the rising MFR. It has been suggested that the occurrence of a CME is determined by the decay index of the strapping field of the MFR, which is derived from the torus instability (TI) of a theoretical MFR model~\citep{kliem06}.

On the other hand, some models consider that the reconnections can happen without the necessity of a pre-existing MFR. In the well-known tether-cutting model~\citep{moore80,moore01}, the preflare magnetic field are strongly sheared, and photospheric converging flows or flux cancellation can push the sheared fields on opposite sides of the neutral line towards one another and results in reconnection of them.
If the reconnection becomes runaway, that is, a feedback between the reconnection and upward expansion of the reconnected long field lines is triggerred,  a flare will be resulted. Then an MFR will form through the continuous reconnection. The MFR can erupt to a CME, or it is confined by a strong overlying field. In another commonly-invoked model, known as the magnetic breakout model~\citep{antiochos99}, it is proposed that reconnection is triggered above the sheared arcades, which is embedded in a quadrupolar configuration that contains a magnetic null point. The shearing of the inner arcade increases its magnetic pressure, making the arcade to expand, which will press the null point and trigger the reconnection. The reconnection weakens the constraint of upper magnetic loops, and trigger the eruption through a feedback between the expansion of the inner arcade and null-point reconnection.


Unravelling what determines the condition for eruptive or non-eruptive (i.e., confine) flare is critical in understanding the mechanism of CMEs. Attempts have been made in observational studies. For instance, \citet{sun15} suggests that some relative measure of magnetic non-potentiality may determine the eruptiveness of active region. \citet{liu16a} suggests that the existence of CME seeds, e.g., some sheared or twisted core field, and the weak enough constraint of background magnetic field are required for producing an eruption.
Numerical magnetodydrodynamic simulations have also been performed to investigate this question. \citet{torok05} simulated two cases of evoluion of an kink-unstable MFR. One successfully erupts while the other is confined because of a stronger overlying field. \citet{devore08} also showed a set of simulations of confined filament/sigmoid. They found that the erupting filaments are decelerated by the background magnetic field in high corona. It seems that the overlying field plays a key role in determining whether CMEs will be formed. Furthermore, in a simulation of the formation and eruption of a sigmoidal MFR by \citet{aulanier10}, it is indicated that reconnection alone is unlikely able to produce the eruption. The MFR can only erupt when it reaches the height to trigger the TI.

In this paper, we investigated an interesting X-class flare event, in which all the three structures as proposed in the theoretical models, namely an MFR, a tether-cutting configuration and a magnetic null point, are involved, while the flare is still confined. This flare of X2.2 class occurred in AR NOAA~12673 with just two hours before the well-known X9.3 eruptive flare in the same region~\citep[the largest flare in solar cycle 24, see][]{yang17,li18,yan18,hou18}. Observations show that there are two episodes of flare brightening, but without eruption. With a magnetic analysis, it is found that during the flare magnetic reconnection is triggered first at the null point aside of the MFR, then followed by a tether-cutting reconnection below the rope, while the rope is firmly confined by its overlying flux. By comparing the magnetic configuration of this flare with that of the X9.3 flare, important insight can be gained into understanding the key factor determining the eruptiveness of flares. The rest of the paper is organized as follows. Data and methods are presented in~Section~\ref{sec:obs}, then the results are given in Sections~\ref{sec:res} and finally conclusions are made in Section~\ref{sec:con}.



\begin{figure*}
\centering
\includegraphics[width=\textwidth]{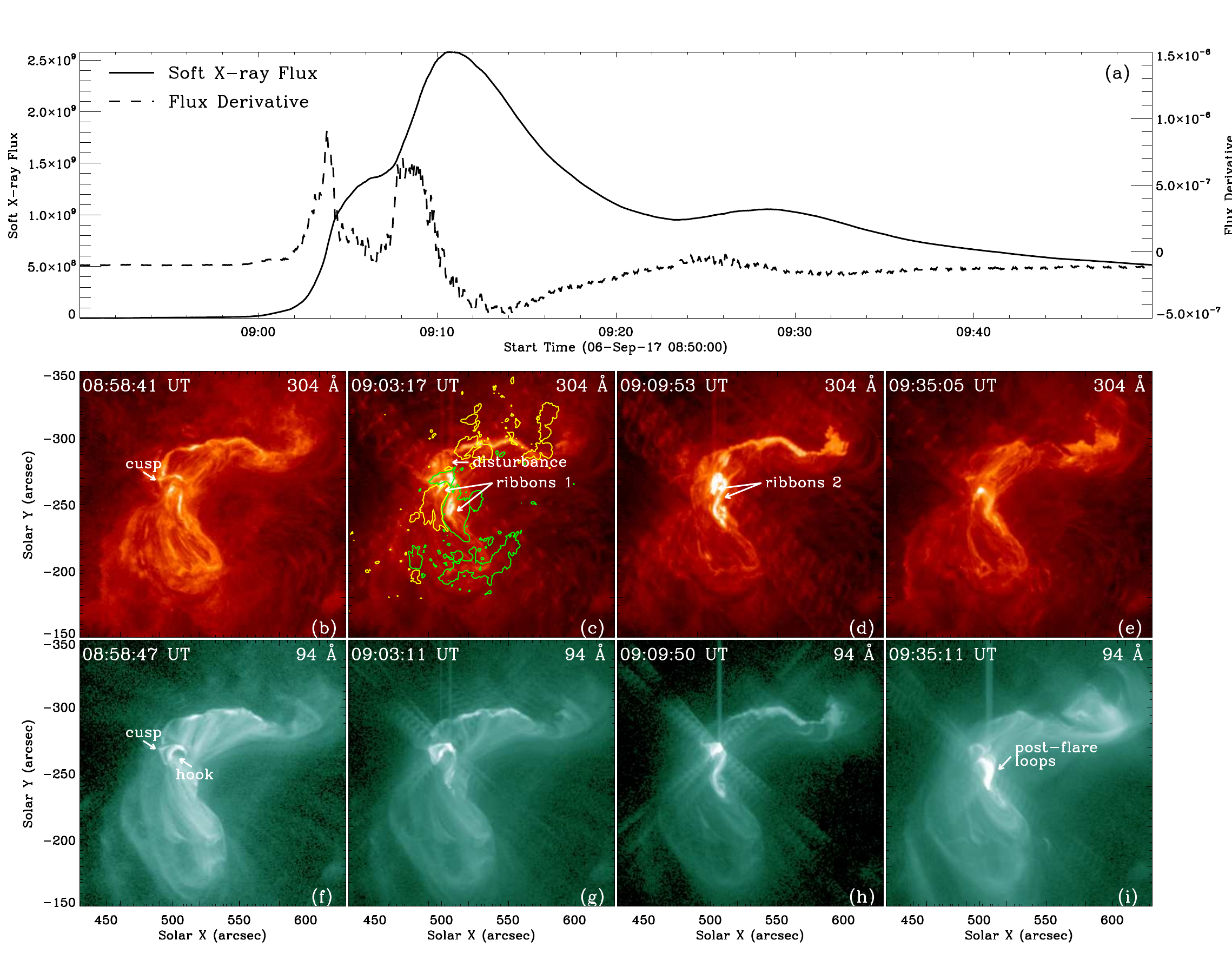}
\caption{The observed evolution of the X2.2 flare. (a) GOES Soft X-ray flux (solid line) and its time derivate (dashed line). (b)-(e) AIA 304 \AA\ images. (f)-(i) 94 \AA\ images. The green (yellow) contour lines in (c) show the line-of-sight photosphere magnetic field with value of $500$~G ($-500$~G). The full evolution of this flare can also be seen in the attached movie in three AIA channels (304, 171 and 94~{\AA}).}
\label{f1}
\end{figure*}

\section{Data and Methods} \label{sec:obs}

For obtaining the full temporal and spatial structures of the X2.2 flare, we used the extreme ultraviolet (EUV) images taken by the Atmospheric Imaging Assembly \citep[AIA,][]{lemen12} onboard the Solar Dynamics Observatory (SDO). It provides 7 EUV filtergrams with a spatial resolution of 0 \farcs 6 per pixel and a cadence of 12s simultaneously. The magnetogram, which can help us understanding the underlying physical process, were taken by the Helioseismic and Magnetic Imager \citep[HMI,][]{sch12,schou12} also onboard SDO. Since the 3D coronal magnetic field can hardly been measured, we extrapolated it from the HMI vector magnetogram, in particular, the Space-weather HMI Active Region Patches \citep[SHARPs,][]{bobra14} dataset. Furthermore, the photospheric motion, which is closely related to the evolution of coronal magnetic field, are derived from the HMI continuum map, which has a spatial resolution of 0 \farcs 5 and cadence of 45s, using Fourier local correlation tracking (FLCT) method~\citep{welsch04}. Finally, the soft X-ray (SXR) flux gained from GOES-13 satellite is used in our study in order to determine the temporal evolution of the flare.

The coronal magnetic field is extrapolated based on the nonlinear force free field (NLFFF) model using the CESE--MHD--NLFFF code developed by~\cite{jiang13}, which is an MHD-relaxation method. It solves a set of modified MHD equations in zero-$\beta$ environment with a friction force. It use an advanced conservation-element/solution-element (CESE) space-time scheme on a non-uniform grid with parallel computing \citep{jiang10}. It was well tested by different benchmarks, such as the analytic force-free solutions \citep{low90} and numerical MFR models \citep{titov99,van04}. Its applications to the SDO/HMI vector magnetograms enable to reproduce magnetic configurations in very good agreement with corresponding observable features, including coronal loops, sigmoids and filaments \citep{jiang13,jiang14}.

The magnetic field data is then analyzed through calculating the magnetic twist number $T_w$, magnetic field decay index $n$, and magnetic squashing factor $Q$. The twist number $T_w$, which is defined by
\begin{equation}
T_{w}=\int_{L}\frac{\left(\nabla\times\textbf{B}\right)\cdot\textbf{B}}{4\pi\textit{B}^{2}}dl,
\end{equation}
quantifies the winding turns between two infinitesimally close field lines \citep{liu16b}. The decay index $n$, which is calculated by $n=-\partial (\log B)/\partial (\log h)$, describe the decaying speed of the strapping field strength $B$ with distance $h$ from the bottom surface. Here $B$ is approximated by the potential field, specifically, its component perpendicular to the path direction along which we compute $n$. Previous works indicate that the torus instability of the constrained MFR will be trigger when $n > 1.5$ \citep{bateman78,kliem06}. For analyzing the magnetic topology, we calculated the magnetic squashing factor $Q$ \citep{titov02}, which can be used to locate thin layers across which the magnetic field-line mapping changes drastically~\citep{demoulin06}. Such layers are so-called quasi-saparatrix layers (QSLs), where 3D magnetic reconnection is prone to occur.

\begin{figure*}
\centering
\includegraphics[width=0.8\textwidth]{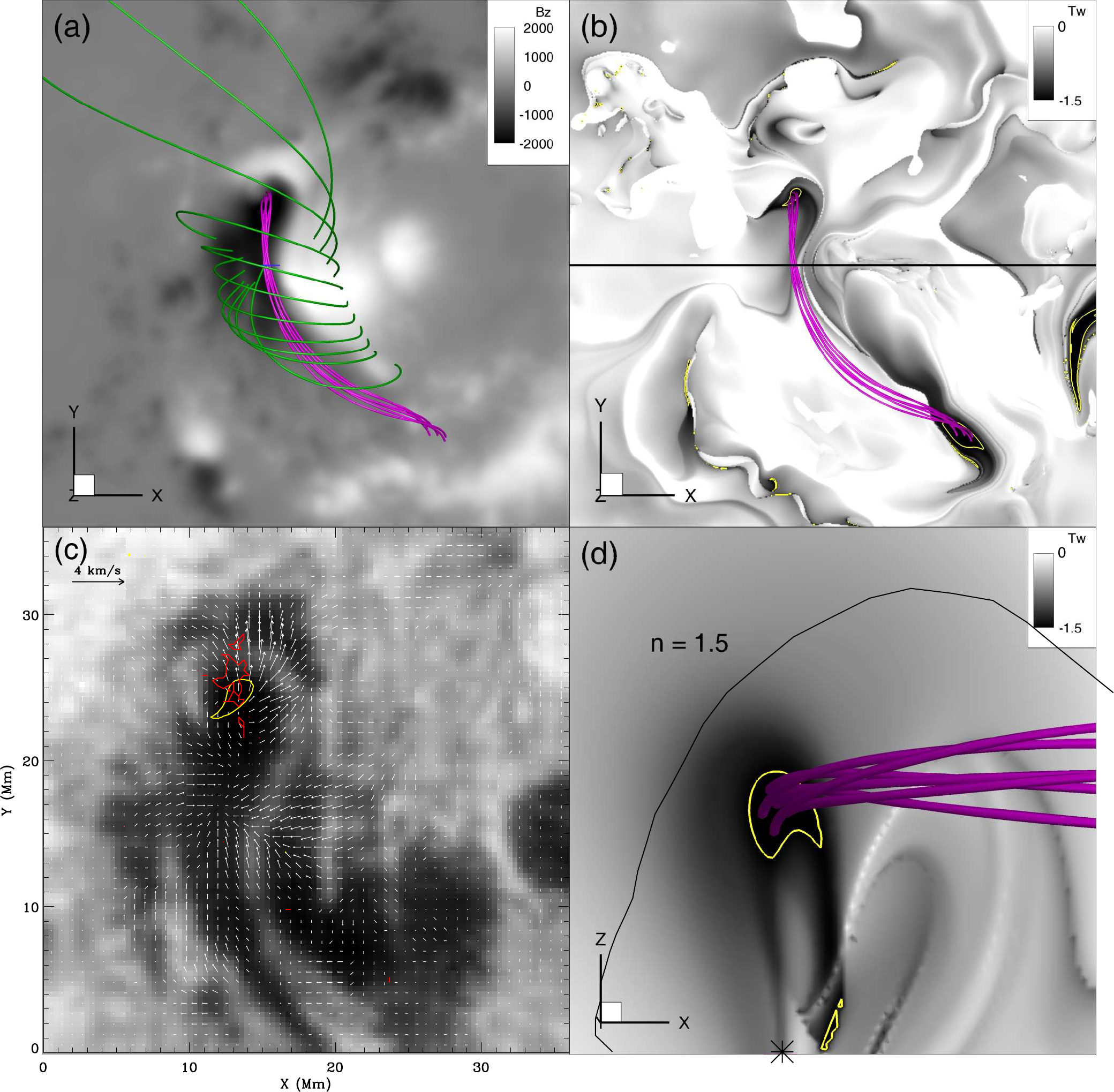}
\caption{Basic configuration of the pre-flare coronal field extrapolated for the time of 08:36~UT. The purple lines in all panels indicate the MFR. In (a), the background is plotted for photospheric $B_z$ map and the green lines are the stapping field lines of the MFR. In (b) the background show the twist number $T_w$ map computed for the bottom surface. (c) The   continuum image overlaid with derived transverse velocity $(v_x, v_y)$ of the photospheric motion. The red line in (c) shows the contour of the area with high curl of the photospheric transverse velocity. (d) A longitudinal cross section and its position is shown in (b) by the thick line. The background is distribution of magnetic twist number $T_w$. The black line denotes the position where the decay index of the strapping field is $n=1.5$. In all panels the yellow curves are contours of $T_{w}=1.5$.}
\label{f2}
\end{figure*}

\section{Results} \label{sec:res}

\subsection{Observations} \label{overview}

Figure~\ref{f1} displays SXR flux and AIA observations (in channels 304 and 94~{\AA}) of the X2.2 flare. The SXR light curve shows that this flare starts at 09:01 UT and peaks at 09:10 UT. Interestingly, there is a short bump of the light curve between the beginning and peak of the flare, and the derivatie of the SXR flux exhibits two peaks, one at 09:04~UT, and the other at 09:08~UT. Such a two-peak feature indicates that there are two eposides of magnetic reconnection during the flare. Indeed, from the AIA observations, it can be found that at 08:58~UT a cusp in the north of the flare core becomes bright (see Figure~\ref{f2}b and f), along with some small-scale material ejecting outward. Immediately, a wave-like disturbance, as can be seen in 304 \AA\ image, propagates outwards from the bright cusp. Then the first flare ribbons formed (denoted as ribbons~1 in Figure \ref{f1}c). The left part of ribbons~1 is shorter and brighter than the right one. With the wave-like disturbance propagating to the end of a hook-like bright loop, the second set of flare ribbons, which is much brighter than ribbons~1, are observed (denoted as ribbons~2 in Figure~\ref{f1}d). The left part of ribbons 2, which is still short and bright, seems to be closer to the PIL than that of ribbons 1, while its right part extends to the north of that of ribbons~1. It means these two sets of flare ribbons, although close to each other, are associated with different reconnection events. Furthermore, there is no separation motion of the flare ribbons between each other. With the brightening of ribbons~2, an inverse S-shaped hot loop is seen in the AIA 94 \AA\ channel (Figure~\ref{f1}h). About 30 minutes after the flare onset, post-flare loops can obviously be seen in the same channel (Figure~\ref{f1}i). From observation of AIA, there is no eruption can be seen accompanied with this flare. Also from the SOHO/LASCO observation, CME is not detected during this flare~\citep{LiuL18}.

\begin{figure*}
\centering
\includegraphics[width=0.8\textwidth]{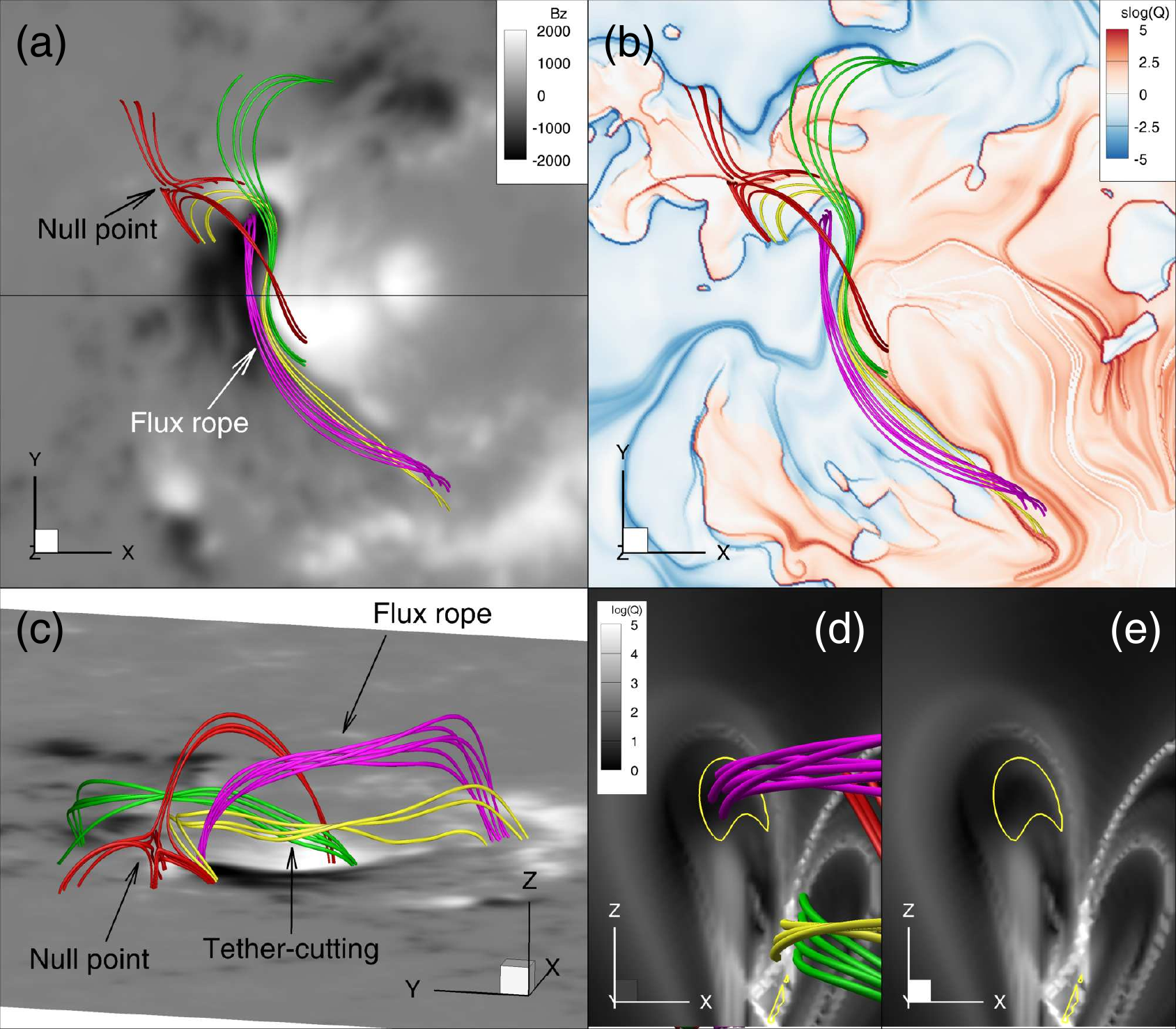}
\caption{Different views of the structures taking part in flare process. The magnetic field is the same extrapolated one shown in Figure~\ref{f2}.
In all the panels, the purple lines are the same set of field lines shown in Figure~\ref{f2}, representing the MFR. Red lines outline the fan-spine configuration around the null point. Note that they are partly overlying the MFR. The yellow lines and green lines are two bundles of magnetic lines forming a configuration ready for tether-cutting reconnection. The background in (a) is plotted for photospheric $B_z$ map, while in (b) it is shown for the magnetic squashing factor $Q$. (c) 3D perspective view of the same structures shown in (a). Panels (d) and (e) are the longitudinal cross sections located at the same position with Figure~\ref{f2}(d), but here both of the backgrounds are plotted for $Q$ map. In (d), the field lines are shown. Note that the yellow and green lines are close to each other with a QSL between them. The yellow contours lines are plotted for the twist degree $T_{w}=1.5$.}
\label{f3}
\end{figure*}

\begin{figure*}
\centering
\includegraphics[width=0.8\textwidth]{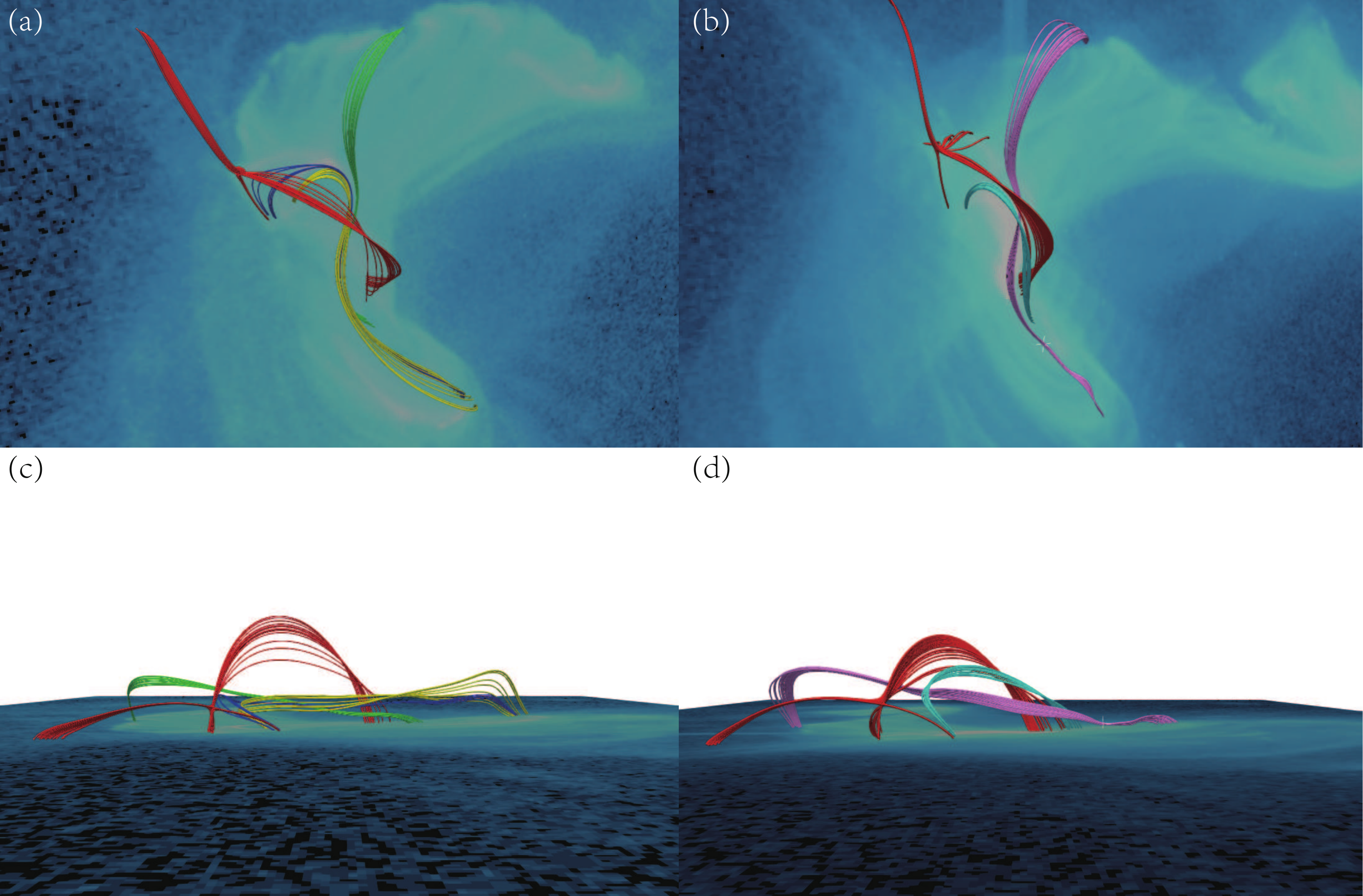}
\caption{The comparison between AIA 94 \AA\ images and extrapolated structures, for the pre-flare phase (panel a) and the post-flare phase (panel b). In the left panel, we use the same colors to indicate the magnetic structures as shown in Figure~\ref{f3}, and the blue lines show the magnetic lines the wave-like disturbance propagating along. In the right panel, the magnetic field is extrapolated for the post-flare state at time 09:36~UT. The cyan lines represent the newly formed post-flare loops, the purple lines show the newly magnetic lines connected two far-end polarities and the red lines for the post-flare null-point configuration. Panel c and d are a side view of panel a and b, respectively.}
\label{f4}
\end{figure*}

\begin{figure*}
\centering
\includegraphics[width=0.8\textwidth]{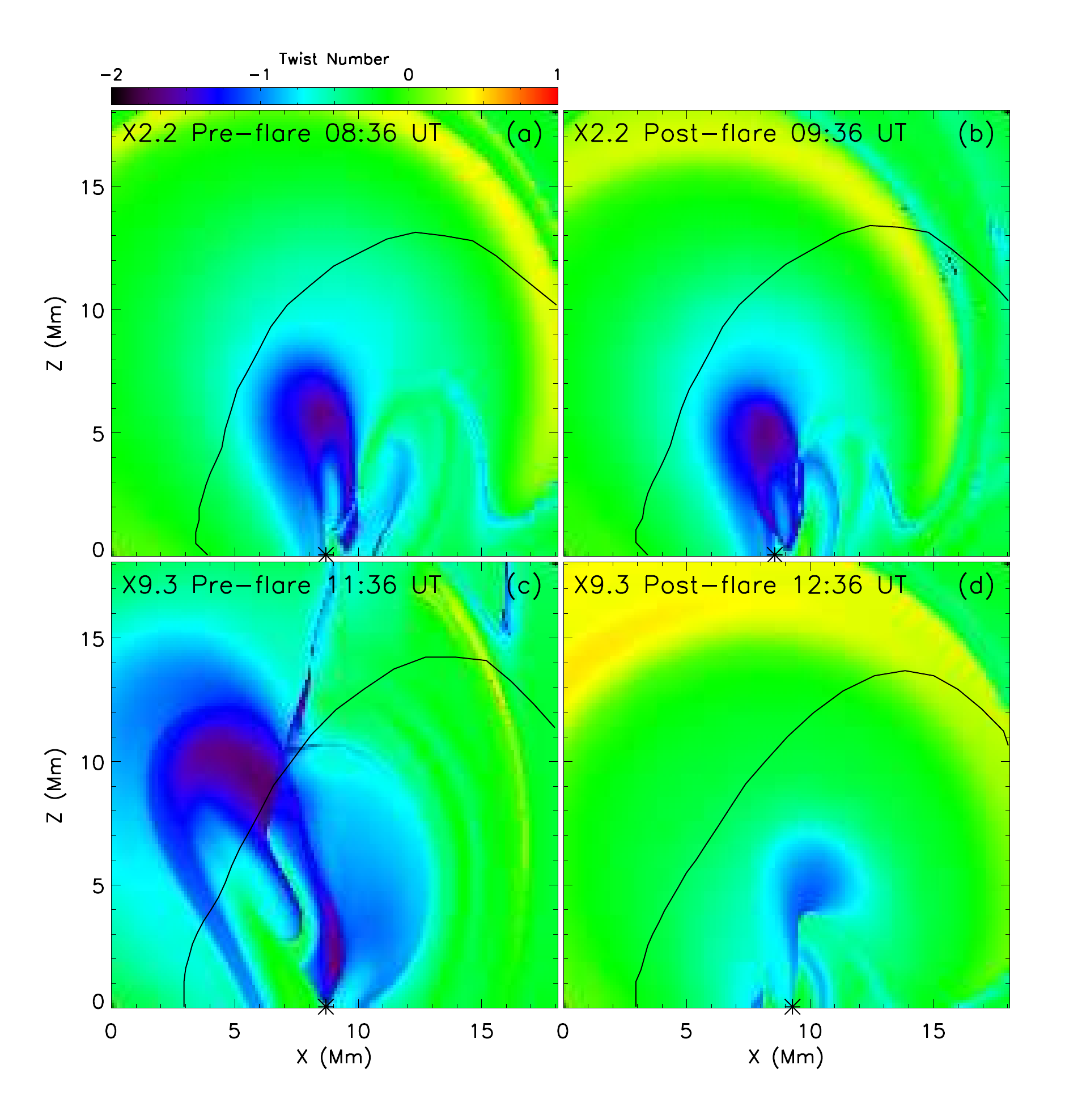}
\caption{Distributions of magnetic twist number on the longitudinal cross section for the X2.2 and X9.3 flares. The magnetic field for the X9.3 flare, onset at 11:53~UT, are extrapolated for the pre-flare time at 11:36~UT and the post-flare time at 12:36~UT.  The position of the cross section is same one shown in Figure~\ref{f2}d.  Top panels are for the X2.2 class flare and bottom panels are for the X9.3 class flare. Left panels for pre-flare phase and right panels for post-flare phase. The black lines in all panels are the contour line of decay index $n = 1.5$, i.e., the threshold of TI.}
\label{f5}
\end{figure*}

\subsection{Magnetic field configuration}
\label{mfc}

Figure~\ref{f2} shows the basic configuration of the pre-flare coronal field, reconstructed for time of 08:36~UT. From the distribution of magnetic twist number, it can be seen that there is a set of twisted field lines with more than one and a half turns ($T_w <= -1.5$), which constitues an MFR lying rougly along the main PIL. The northern (negative-polarity) foot of the MFR is found to be coincide with an anticlockwise rotating sunspot~\citep[see also in][]{yan18}, and particularly, the area of strongest twist is co-spatial with the strongest vorticity of the photospheric surface motion. This provides a strong evidence that the rotating sunspot built and was further strengthing the MFR. In Figure~\ref{f2}d, we show the decay index $n$ of the overlying field of the MFR at a central cross section of the volume. It can be seen that the rope is fully below the line with $n=1.5$, which is found to be a typical threshold of TI in theoretical and numerical calculations for a line-tied arched MFR~\citep{aulanier10,zuccarello16}. This suggests that the rope is firmly confined by its overlying field, which can explain why the flare is not eruptive, or, at least, this MFR did not erupt during the flare. Indeed, by comparing the pre-flare and post-flare magnetic field, there is very limited variation of the magnetic twist distribution (to be discussed in Section~\ref{compare}).

Then, what causes this flare? Analysis of the magnetic topology provides important insight. Figure~\ref{f3} shows the basic building blocks of the magnetic topology of the pre-flare field. There exists a null-point configuration in the north of the MFR, and the field lines constituting spine-fan configuration of the null extends overlying the rope (see the red lines in Figure~\ref{f3}), although the null itself does not overlie the rope. Furthermore, there is a QSL with squashing factor $Q>10^5$ right below the rope, and field lines threading this QSL forms the typical configuration in the tether-cutting model (see the yellow and green lines in Figure~\ref{f3}). One set of these field lines (the green ones) connects to the northwest, and the other set (colored in yellow) forms a hook shape around the MFR's northern foot, extending close to the null point. As these field lines, as well as the null-point configuration, are prone to take part in reconnection, in Figure~\ref{f4}(a) we show them in the same view angle of the SDO and overlaid on the AIA 94 \AA\ image. A nice match of the field lines with the loops can be seen. As mentioned in section~\ref{overview}, the flare first brights at a cusp-like configuration, and when compared with the magnetic field lines, it is evident that this cusp corresponds to the null-point structure, and reconnection here produced the flare ribbons~1. Location of flare ribbons~2 are also in good agreement with the footpoints of the tether-cutting configuration. Particularly, the field lines colored in yellow (see Figures~\ref{f3} and \ref{f4}(a)), whose north footpoints are rooted in the left flare ribbon, are well co-spatial with the aforementioned hook-shaped loops. Also, the south footpoints of the green lines are rooted in the right flare ribbon. The inverse S-shaped hot loop in Figure~\ref{f1}h is a result of this tether-cutting reconnection.

From the above analysis, we suggest an scenario for this flare process. The rotating sunspot twists the MFR at its footpoint, thus the MFR expands~\citep{yan18} because of increasing magnetic pressure. Meanwhile, the sunspot moves northward, also stressing the MFR. Both the expansion of the MFR and the stressing motion can push magnetic flux to the null point, leading to current sheet forming there, and finally triggers reconnection when coronal resistivity takes effect in the current sheet. Thus the null-point reconnection results in the cusp-shaped brightening. Meanwhile, this null-point reconnection will release the magnetic tension force that confines the MFR (but with only a little bit because the post-reconnected field is still closed rather than open). The MFR then quickly expands upward a little following the null-point reconnection. Consequently, magnetic pressure below the rope is weakened, which can trigger the tether-cutting reconnection. Also the wave-like disturbance from the null-point reconnection propagates along the field lines (see the blue one in Figure~\ref{f4}a) can perturbate the field underlying the rope, which also might provide a trigger for reconnection there. The tether-cutting reconnection results a short, inverse S-shaped post-flare loop (see the field lines colored in cyan in Figure~\ref{f4}b) and a long field lines connecting the two far-end polarities in south and north (the purple one) respectively. We note that the null-point configuration still exists after the flare as suggested by the extrapolation for post-flare field. As can be seen in Figure~\ref{f4}c and d, the null point was lifted up slightly from $\sim 4$ Mm to $\sim 6$ Mm.

\subsection{Comparison with the X9.3 flare}
\label{compare}
In the same region of this confined flare, after only $\sim 2$ hours, the X9.3 flare occurred with a large eruption and a CME~\citep{yan18}. With such a short time interval, the basic magnetic configuration should be similar for these two flares. Thus, it is important to understand what is the key factor that determines the eruptiveness. We extrapolate the magnetic fields of the X9.3 flare as well and in Figure~\ref{f5} we compare the distributions of magnetic twist $T_w$ and decay index $n$ in the central cross section, from pre-flare to post-flare state, for the two flares. As can be seen, both pre-flare states have magetic flux of high twists of around 1.5 turns, which form the pre-flare MFRs. Furthermore, the locations of decay index $n=1.5$ (i.e., the theoretical TI threshold line) are similar for both pre-flare states, suggesting that the strapping flux configuration is similar.
In contrast, the pre-flare locations of the MFR differ considerably. For the X2.2 flare, the MFR is fully underneath the TI threshold line (Figure~\ref{f5}a), while for the X9.3 flare, the major part of the MFR is above the threshold line (Figure~\ref{f5}c). For the post-flare states, the high-twist flux, i.e., the MFR, of the X2.2 flare still exists and stays in the position similar to its pre-flare state (see panel b), but the MFR of X9.3 flare disappeared because of the eruption (see panel d).

Interestingly, by comparing Figure~\ref{f5}a and b, the post-flare MFR is at a slightly lower height than that of the pre-flare one, which seems to be contradictory to our scenario, i.e., the MFR is supposed to be lifted up. This is probably due to the limitation of NLFFF extrapolation. The AIA observation indicates that the post-flare loops are still forming and evolving, indicating that actually the coronal magnetic field is still rather dynamic, which cannot be modeled by the static extrapolation. Thus, in a dynamic state, the flux rope could be higher than the extrapolated one.

\section{Conclusion} \label{sec:con}

In this paper, we studied an interesting confined flare of X2.2 class that took place in AR~12673. It exhibits rather complex flare ribbons with two episodes of brightening, while no eruption is observed. For understanding the underlying magnetic process, we extrapolated the coronal magnetic field using an NLFFF model, and by analyzing the magnetic topology, it is revealed that three important magnetic structures are involved within the flare. They are, respectively, an well-defined MFR, a magnetic null point, and a tether-cutting configuration, that is, two sets of magnetic field lines ready for tether-cutting reconnection. The null situates close to the northern leg of the MFR, with part of its spine-fan structure overlying the MFR. The tether-cutting field lines runs below the MFR, forming a QSL there. One set of the tether-cutting field lines also extends close to the null point, forming a hook around the northern leg of the MFR. In particular, the MFR's northern leg is found to be rooted in a rotating sunspot moving toward to the direction of the null point. Previous studies have found that such photospheric motions play an important role in transporting energy and triggering reconnection for the X-class flares occurred in this AR~\citep{romano18, verma18}.
From the complex magnetic configurations and the photospheric motions, we suggest that the rotating sunspot drives the MFR to expand, which pushes magnetic flux to the null point, and reconnection is first triggered there. Then the disturbance from the null-point reconnection as well as the subsequent upward expansion of the MFR trigger the second reconnection, i.e., the tether-cutting reconnection. Such a scenario is compatible with the AIA observations, as the modeled field lines show a good match with the observed loops and their footpoints match the flare ribbons very well. Interestingly, neither the null-point reconnection nor the tether-cutting process results an eruption. The key factor that determines the confinement of the flare is found to be the MFR's strapping field, since by calculating its decay index, we indicate that the strapping flux can firmly hold the MFR during the flare. Furthermore, we compared the configuration of this flare with that of the eruptive X9.3 flare just 2 hours later, which has a similar MFR configuration. The key difference between them is that, for the confined flare, the MFR is fully below the TI threshold, while for the eruptive one, the main body of MFR reaches above the threshold.

In summary, we observed a confined X-class flare and provided a scenario for the magnetic mechanism of the flare through a sophisticated analysis of extrapolated coronal magnetic field. Although the magnetic configuration is complex with different topological structures involved in the flare, the determining factor of the confinement is the strapping field that stabilizing the MFR or the eruptive current system in the AR's core. The study of this flare provides a good evidence supporting that reconnection might not be sufficient to trigger eruption, in which MHD instability plays a more important role.

\acknowledgments
This work is supported by the National Natural Science Foundation of China (NSFC 41822404, 41731067, 41574170, 41531073), the Fundamental Research Funds for the Central Universities (Grant No. HIT.BRETIV.201901). Data from observations are courtesy of NASA {SDO}/AIA and HMI science teams.

\end{document}